%% file: main.tex
\documentclass[preprint]{vgtc}               

\ifpdf
  \pdfoutput=1\relax                   
  \usepackage{graphicx}                
  \DeclareGraphicsExtensions{.pdf,.png,.jpg,.jpeg} 
\else
  \usepackage{graphicx}                
  \DeclareGraphicsExtensions{.eps}     
\fi%

\graphicspath{{figures/}{pictures/}{images/}{./}} 

\usepackage{microtype}                 
\PassOptionsToPackage{warn}{textcomp}  
\usepackage{textcomp}                  
\usepackage{mathptmx}                  
\usepackage{times}                     
\usepackage{cite}                      
\usepackage{tabu}                      
\usepackage{booktabs}                  

\usepackage{url}
\usepackage{graphicx}
\usepackage{csquotes}
\usepackage{listings}
\usepackage{color}
\definecolor{lightgray}{rgb}{.9,.9,.9}
\definecolor{darkgray}{rgb}{.4,.4,.4}
\definecolor{purple}{rgb}{0.65, 0.12, 0.82}
\lstdefinelanguage{JavaScript}{
  keywords={break, case, catch, continue, debugger, default, delete, do, else, false, finally, for, function, if, in, instanceof, new, null, return, switch, this, throw, true, try, typeof, var, void, while, with},
  morecomment=[l]{//},
  morecomment=[s]{/*}{*/},
  morestring=[b]',
  morestring=[b]",
  ndkeywords={class, export, boolean, throw, implements, import, this},
  keywordstyle=\color{blue}\bfseries,
  ndkeywordstyle=\color{darkgray}\bfseries,
  identifierstyle=\color{black},
  commentstyle=\color{purple}\ttfamily,
  stringstyle=\color{red}\ttfamily,
  sensitive=true
}

\usepackage[normalem]{ulem}

\newif\ifnotes
\notesfalse 

\definecolor{purplecolor}{RGB}{128, 0, 128}
\definecolor{grey}{RGB}{128, 128, 128}

\DeclareRobustCommand{\hsout}[1]{\texorpdfstring{\renewcommand{\cite}{\ccite}\sout{#1}}{#1}}
\DeclareRobustCommand{\del}[1]{{\ifnotes{\leavevmode\color{grey}{\protect\hsout{#1}}}\fi}}
\newcommand{\add}[1]{\ifnotes{\color{purplecolor}{#1}}\else{#1}\fi}
\newcommand{\replace}[2]{\ifnotes{\del{#1}\add{#2}}\else{#2}\fi}

\onlineid{1086}

\vgtccategory{Research}

\vgtcinsertpkg

\title{Representing Real-Time Multi-User Collaboration in Visualizations}

\author{Rupayan Neogy\thanks{e-mail: rneogy@mit.edu}\\ \scriptsize MIT CSAIL\vspace{-5mm} %
\and Jonathan Zong\thanks{e-mail: jzong@mit.edu} \\ \scriptsize MIT CSAIL\vspace{-5mm}
\and Arvind Satyanarayan\thanks{e-mail: arvindsatya@mit.edu} \\ \scriptsize MIT CSAIL\vspace{-5mm}}

\input{sections/abstract}

\CCScatlist{
  \CCScatTwelve{Human-centered computing}{Visu\-al\-iza\-tion}{Visu\-al\-iza\-tion systems and tools}{};
}

\begin{document}


\firstsection{Introduction}

\maketitle


\input{sections/intro}
\input{sections/related_work}
\input{sections/design_space}

\input{sections/evaluation}
\input{sections/conclusion}

\acknowledgments{
We thank our anonymous reviewers for their constructive feedback. 
This work was supported by NSF Award \#1942659.
}

\bibliographystyle{abbrv-doi}

\bibliography{main}
\end{document}

%% file: sections/abstract.tex
\abstract{
Establishing common ground and maintaining shared awareness amongst participants is a key challenge in collaborative visualization.
For real-time collaboration, existing work has primarily focused on synchronizing constituent visualizations\,---\,an approach that makes it difficult for users to work independently, or selectively attend to their collaborators' activity.
To address this gap, we introduce a design space for representing synchronous multi-user collaboration in visualizations defined by two orthogonal axes: \emph{situatedness}, or whether collaborators' interactions are overlaid on or shown outside of a user's view, and \emph{specificity}, or whether collaborators are depicted through abstract, generic representations or through specific means customized for the given visualization.
We populate this design space with a variety of examples including generic and custom synchronized \emph{cursors}, and \emph{user legends} that collect these cursors together or reproduce collaborators' views as thumbnails.
To build common ground, users can interact with these representations by \emph{peeking} to take a quick look at a collaborator's view, \emph{tracking} to follow along with a collaborator in real-time, and \emph{forking} to independently explore the visualization based on a collaborator's work.
\replace{We evaluate our contribution via}{We present a reference implementation of} a wrapper library that converts interactive Vega-Lite charts into collaborative visualizations\replace{, and}{. We} find that our approach affords synchronous collaboration across an expressive range of visual designs and interaction techniques.

} 

%% file: sections/intro.tex
As interactive visualization is increasingly integrated into data journalism, science communication, and public health response, there is a growing need for research on how interactivity might facilitate collaborative sensemaking.
To meet these changing needs, a recent thread of research in \textit{collaborative visualization} has investigated methods for multiple users to interactively explore and analyze the same visualization across multiple locations and devices.
Systems such as Polychrome~\cite{polychrome} and Vistrates~\cite{vistrates} provide techniques for synchronizing a visualization, such that interactions performed on one device immediately appear on all others. Though these systems recognize the value of maintaining shared awareness of collaborators' interactive exploration, they do not yet support ways of doing so while allowing users to conduct their own independent, divergent exploration.
\del{Systems such as} sense.us~\cite{2007-senseus} and Many Eyes~\cite{manyeyes} support collaborative analysis through graphical annotations and textual comments, and information scent techniques have also been explored to depict the expectations of other readers~\cite{2018-others-expectations}.
However, such approaches target \emph{asynchronous} collaboration;
\replace{Thus, methods for unobtrusively representing multiple users synchronously collaborating around a shared visualization remain under-explored.}{for synchronous collaboration, prior work has considered only specific interaction techniques such as brushing \& linking~\cite{hajizadeh2013supporting} or query widgets~\cite{hong2018collaborative}.}

To \replace{address this gap}{generalize from these approaches}, we introduce a design space for augmenting visualizations with representations of remote users that enable shared awareness without sacrificing independent exploration. 
This design space is described by two orthogonal axes: \replace{\textit{situatedness}, or whether collaborators’ interaction are overlaid on or shown outside of a user’s view, and \textit{specificity}, or whether collaborators are depicted through abstract, generic representations or through specific means customized for the given visualization.}{\textit{situatedness}, or whether collaborators are represented within the visualization or externally, and \textit{specificity}, or whether collaborators' representations are generic or tailored to the visual idioms of the visualization.} We present designs for \textit{cursors}, high-level representations of remote users that encode information about their interactions. Cursors do not actively modify the visualization state, allowing users to maintain independent viewing experiences, while providing \textit{information scent} \cite{information-scent} of remote interaction.
By default, cursors appear as small color-coded dots on the visualization, situated in the location of a remote user's interaction. However, their appearance can be customized to increase their specificity to a particular visualization.
We also introduce designs for \textit{user legends}, which are collections of user representations situated alongside the visualization. Legends can similarly be more or less specific to a visualization. For instance, legends may collect cursors into one place to prevent crowding within the visualization, or fully represent remote user views as thumbnails.

In addition to these visual approaches, we contribute a set of interactive methods to help users build common ground.
These techniques include \textit{peeking}, a temporary look at a remote user’s view; \textit{tracking}, a continuous synchronized view of a remote user's visualization; and \emph{forking}, a divergent exploration starting from a remote user's state. By interacting with cursors, users can examine their collaborators' interactive exploration in an ambient fashion. For example, a user can hover over a cursor to peek at the remote user's view before returning to their previous interaction state, or they can click on the cursor to actively update their visualization state with the remote one. In this way, each user has their own visualization to explore, and remote interactions do not affect their experience until the user chooses to interact with a cursor. Augmenting each user's separate instance ensures that users gain collaborative benefits while retaining agency over their own exploration.

\replace{To evaluate}{To validate} our designs, we implement \del{these techniques in }a wrapper for the Vega-Lite visualization grammar~\cite{vegalite} that converts Vega-Lite specifications into synchronous, collaborative visualizations. 
We explore a variety of user representation designs, demonstrating that our contribution affords collaboration across an expressive range of visualization designs and interaction techniques.

%% file: sections/related_work.tex
\section{Related Work}\label{chap:related_work}

Our work builds on prior methods for collaborative visualization and computer-supported cooperative work (CSCW), and draws inspiration from techniques for representing multiple users collaborating around textual documents.

\subsection{Collaborative Visualizations and Documents}

Early systems such as Many Eyes~\cite{manyeyes}, the Baby Name Voyager~\cite{wattenberg2005baby}, and sense.us~\cite{2007-senseus} helped identify that visualizations can indeed serve as social artifacts.
Through graphical annotations and textual comments, users of these systems eagerly engaged in collaborative data analysis. 
However, this activity could only occur asynchronously and, in a more recent thread of work, researchers have explored how visualizations can also facilitate synchronous collaboration.
For example, systems like Polychrome~\cite{polychrome} and Vistrates~\cite{vistrates} support both asynchronous and synchronous collaboration by allowing users to keep their visualizations synchronized with one another.

Reflecting on this work, researchers have defined the terms \emph{collaborative visualization}~\cite{collab-vis-definition} and \emph{ubiquitous analytics}~\cite{ubiquitous-analytics} and have identified their associated challenges~\cite{collab-vis} and design considerations~\cite{design-considerations}.
In particular, a chief concern is how to establish common ground among collaborators, and allow them to maintain a shared awareness of each other's work. 
sense.us\del{, for example,} addressed this concern \del{by }through both traditional social navigation mechanisms (e.g., user profiles and comment listings) and by introducing ``doubly-linked'' discussions where textual comments deep-link to specific interactive visualization states, and corresponding textual comments automatically populate the discussion panel as a user interacts with the visualization~\cite{2007-senseus}. 
However, for synchronous collaborative visualization, existing work \replace{has not yet grappled with this concern as it has}{instead} primarily focuse\replace{d}{s} on the backend frameworks and architectures necessary to support this type of behavior~\cite{polychrome, vistrates}.

Here, collaborative word processing packages such as Google Docs, Etherpad, and Overleaf suggest a way forward. 
Like their visualization counterparts, these tools enable synchronous collaboration by maintaining a shared state and keeping each user's view synchronized.
As a result, common ground is established in a straightforward fashion as every user sees the same visual output. 
And, to maintain a shared awareness, these systems augment the view by assigning each user their own color-encoded cursor.
\add{Researchers have adopted these ideas for visual analytics systems but in the context of specific visual idioms or interaction techniques\,---\,for instance, using color-coded representations for brushing \& linking~\cite{hajizadeh2013supporting}, query widgets~\cite{hong2018collaborative}, and externalizations of network data~\cite{mahyar_tory_2014}.}
Our work \del{is inspired by this approach, but motivated by principles from the CSCW literature described below, }goes a step further by \replace{exploring a richer}{generalizing these point examples into a rich} design space \add{that is agnostic to or customized for mark types and is applicable to a broad range of selection-based interaction techniques. Moreover, our work also considers how to instrument these shared awareness representations with techniques for re-establishing common ground.}
\del{In particular, we consider both in situ and external ways of representing remote users, as well as designs that are agnostic to or customized for visualization types.}

\subsection{Computer-Supported Cooperative Work (CSCW)}

The CSCW literature organizes collaborative behavior using the \emph{time-space matrix}: the time dimension indicates whether participants interact synchronously or asynchronously, and the space dimension describes whether collaborators are co-located or geographically distant~\cite{applegate1991technology, johansen1988groupware}.
Wood et al. instantiated these ideas in the space of visualization, coining the term computer supported collaborative visualization (CSCV)~\cite{wood1995cscv}.
In a literature survey~\cite{brodlie2004distributed}, they situate a variety of visualizations systems within this matrix.
However, this discussion is primarily concerned with visualization models and architectures (i.e., the components and abstractions available for constructing distributed or collaborative visualizations) and does not dwell on aspects of user experience.
Our work addresses the synchronous (same time) column of the time-space matrix, and is agnostic to whether users are co-located or distant.

Our work is moreover guided by Grudin's principle of \emph{unobtrusive accessibility}: collaborative features should not disrupt an individual's activity~\cite{grudin1994groupware}.
This principle helps us identify a shortcoming with simply synchronizing views across participants.
While only mildly disruptive for collaborative document editing, keeping views synchronized can prevent independent exploration of visualizations for a large swath of interactions performed by remote users\,---\,for example, filtering data out or panning and zooming to explore a different subset of the data.
In contrast, under our approach, each user maintains their own independent instance of a visualization and interactive state is synchronized and depicted through in situ or external graphical representations (\autoref{fig:overview}).
To mitigate any loss of common ground, we additionally contribute interactive behaviors that a user can use to \emph{peek} into or \emph{track} their collaborators' activity.

%% file: sections/design_space.tex
\section{A Design Space for Representing Remote Users}\label{chap:design}

\add{
To formalize the design dimensions for representing remote users on visualizations, we used an iterative design process seeded by prior approaches discussed in the previous section.
In particular, we sought to distill a minimal set of dimensions capable of covering a diverse range of visualization types and interaction techniques.
We prototyped candidate designs using Vega-Lite~\cite{vegalite}, as its expressive grammar allowed us to systematically enumerate mark types and interaction techniques.
To refine our designs, we conducted informal feasibility tests with members of our research group and other students at our research institution.
These tests revealed the tradeoffs at stake: while participants preferred direct annotations on the visualization, this could also cause visual crowding or conflicts with interactions that alter the main view.
We ultimately conceptualize the design space of representations of remote users along the following two orthogonal dimensions.
}

\del{To support our goals of enabling independent exploration, maintaining shared awareness, and establishing common ground, we formulate collaborative visualization as users interacting with their own individual view augmented with
marks representing remote users synchronously. We conceptualize the design space of possible representations along two orthogonal dimensions.}

\textbf{Situatedness} refers to the location of user representations in relation to the visualization: marks representing remote user interactions can be placed either in situ (overlaying the visualization) or externally. 
This axis describes the tradeoff between depicting remote users in context and the degree to which doing so can disrupt reading a visualization.
For instance, in situ representations can leverage the visualization's encodings, scales, axes, and legends to aid a viewer's awareness of their collaborator's activity.
On the other hand, such representations may also crowd the visualization and obscure legibility, especially when there are large numbers of remote users.
In such cases, external representations might be preferred.

\textbf{Specificity} \replace{refers to whether collaborators are depicted through abstract, generic representations or through specific means customized for the given visualization.}{refers to whether user representations are designed generically, without respect to the content of the visualization, or if they are customized to the visual idioms and interaction techniques of a particular visualization.}
For example, representing users with minimal circular marks will generically apply to any visualization without much knowledge of what interactions it supports. However, on a scatterplot with a brushing interaction, we can represent remote users as color-coded, synchronized rectangles\,---\,a representation tailored to this specific interaction technique.
Specificity determines how information-dense representations of remote users are:
a more specific representation communicates more information about a remote user, but requires added effort to design.

\begin{figure}
    \centering
    \includegraphics[width=\columnwidth]{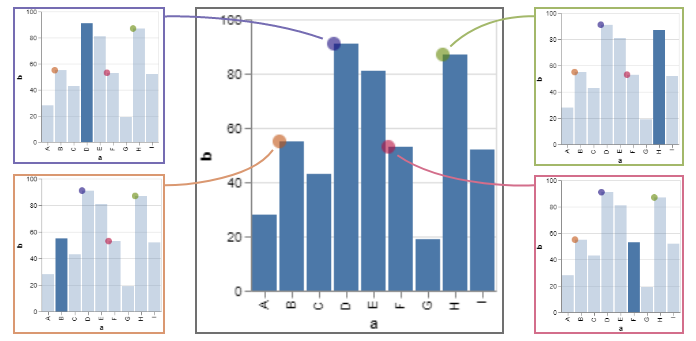}
    \caption{Selections from various remote users are translated into interactive cursors on each view.}
    \label{fig:overview}
    \vspace{-3mm}
\end{figure}

\begin{figure*}
    \centering
    \includegraphics[width=0.99\textwidth]{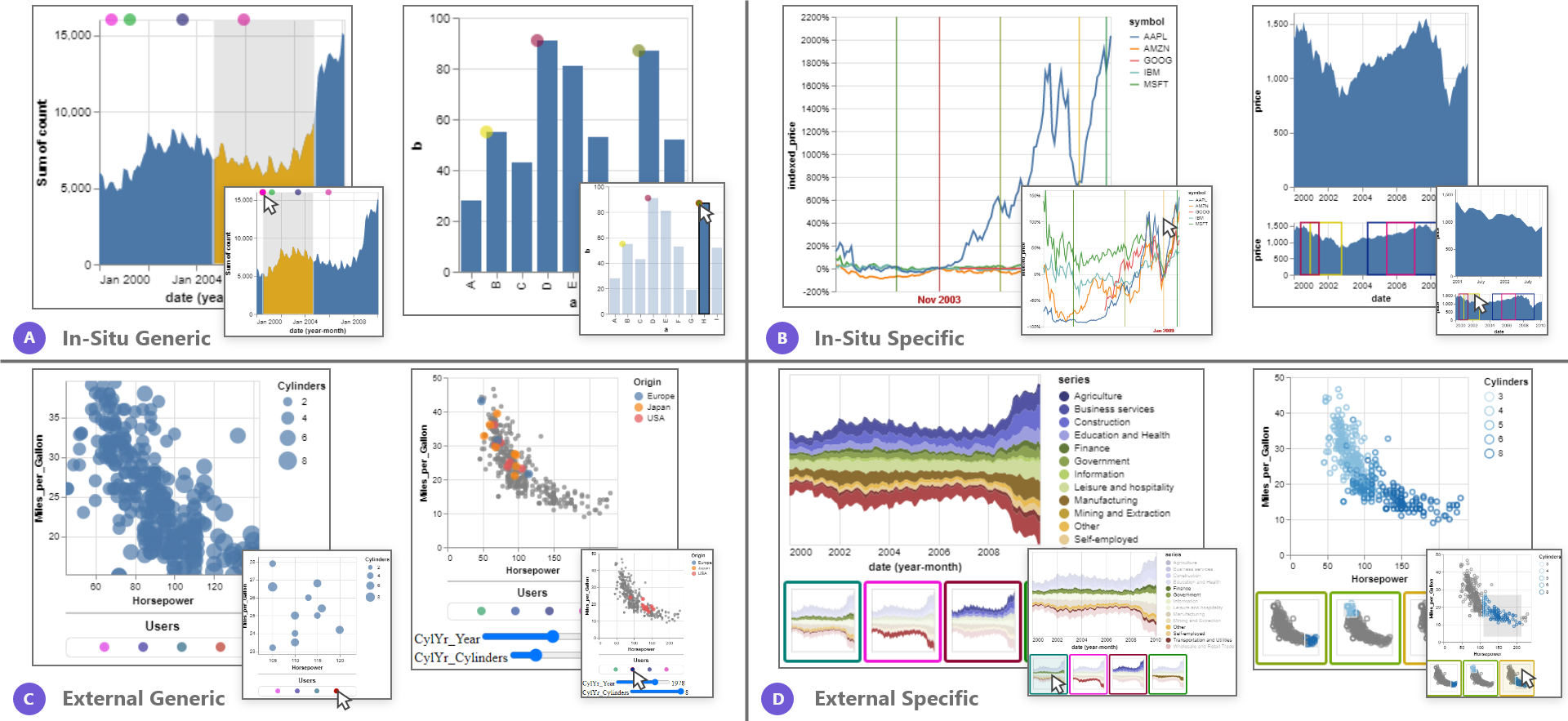}
    \caption{Example gallery of user representations on collaborative visualizations. Inset views show the result of the \textit{peeking} hover interaction for each example. A) Generic cursors overlay each visualization, indicating remote interaction. B) Specific cursors provide more visualization-specific information about remote user interaction. C) Cursor legends provide an external, minimal representation of remote users. D) Thumbnail legends visually communicate a specific overview of the remote interaction state.}
    \label{fig:gallery}
    \vspace{-4mm}
\end{figure*}

We use \textit{cursor} to refer to the basic unit of user representation in a collaborative visualization. Each cursor corresponds to a single remote user and can vary in appearance based on trade-offs between \textit{situatedness} and \textit{specificity}.
\replace{Figure \ref{fig:gallery}}{\autoref{fig:gallery}} provides a variety of example designs that occupy different positions within these two dimensions.

\subsection{Generic Cursors}

Generic cursors, shown in \replace{Figure \ref{fig:gallery}}{\autoref{fig:gallery}}A, are small, translucent, color-coded marks overlaid on the visualization in the position of each remote user's interaction.
They are designed to provide awareness of remote collaborators in context while minimally interfering with a user's visualization experience\,---\,for instance, users can selectively attend to them, and they are sufficiently unobtrusive to be ignored altogether.
Moreover, their generic design allows them to operate across a variety of visualizations (e.g., histograms, scatter plots, line charts, etc.) and interaction techniques (e.g., reflecting a user's mouse position, the top-left coordinate of a brushed region, etc.).


\replace{In our design process, w}{W}e explored various generic abstractions of remote user interactions. Given the rich space of interaction techniques for data visualizations~\cite{heer2012interactive, yi2007toward}, we found position to be the property of remote user behavior that is most broadly applicable and meaningful\,---\,it either depicts attributes of interaction (e.g., brush extents) or falls back to reflecting the remote user's cursor location.


\subsection{Specific Cursors}


In situ cursors can be customized heavily in a variety of ways to present more specific representations of remote user interactions. \replace{Figure \ref{fig:gallery}}{\autoref{fig:gallery}}B illustrates how color-coded vertical rule cursors indicate the index points remote users are exploring when interactively re-normalizing stock price time-series data.
The same figure also presents a synchronized version of an overview\,+\,detail plot, where brushes on the lower chart change the zoom and pan of the upper view. Custom color-coded rectangles outline the brushed regions from remote users.
Note, while our examples use color to distinguish remote users, any categorical encoding channel can be used to do so (e.g., encoding each remote user using a unique shape).

Specific cursors replicate recent approaches for representing users in collaborative visualization (e.g., synchronized brushes and selections \`a la Vistrates~\cite{vistrates}, or zoom windows, as presented in Visualive~\cite{2019-towards-visualive}).
They afford more information-dense displays of remote user interactions but risk trading off readability.
For example, having multiple brush outlines on the overview can interfere with a user's ability to define a new brush. 
Similarly, the vertical lines on the stock-index chart can potentially confuse the viewer about what point in time is rescaling the data. These examples demonstrate that specific cursors must be designed carefully and with purpose to serve a specific collaborative goal, whether that may be presentation or independent exploration of the data.

\subsection{Cursor Legends}

While in situ cursor designs work for a variety of visualizations types and interaction techniques, there are many instances where an external representation may be preferred. 
For example, perhaps the visualization is already very data-dense, such as a scatter plot with a large number of points. Overlaying translucent circular cursors could obfuscate the data or confuse the viewer. 
For cases where overlaying cursors is not appropriate or effective, we present the \emph{cursor legend}, an off-visualization container for cursors as shown in \replace{Figure \ref{fig:gallery}}{\autoref{fig:gallery}}C.
The first example uses a cursor legend to depict remote panning \& zooming behaviors.
This interaction technique presents an interesting challenge: a remote user may have panned or zoomed entirely out of the current user's view.
In such a situation, in situ cursors would either disappear entirely or could be made to clamp to the corners of the view rectangle\,---\,in either case, there is the potential for confusion from their typical behavior.
With a cursor legend, all remote users are presented in a consistent manner regardless of where they have panned or zoomed to.
The second example illustrates dynamic query widgets.
Here too, the cursor legend is arguably a more natural representation both because the interactive widgets are external to the visualization and because a remote user's interactive state is the joint combination of all sliders\,---\,an in situ alternative might overlay each slider with a cursor but would require additional effort from a viewer to piece back together.

In summary, cursor legends preserve the original visualization experience, with no modifications or overlays, but trade off the ambient information-scent affordance of their in situ counterparts.
Moreover, cursor legends can also be used in conjunction with in situ cursors as a secondary representation of remote users, or to provide more information for each user (e.g., usernames or avatars as often seen in collaborative word processing tools).

\subsection{Thumbnail Legends}

Thumbnail legends, \del{as }shown in \replace{Figure \ref{fig:gallery}}{\autoref{fig:gallery}}D, recall bookmark trails from sense.us~\cite{2007-senseus} as well as graphical histories~\cite{2008-graphical-histories}.
They provide \del{simplified, }high-level previews of remote visualizations and their interactive state. 
Just as specific cursors provide an alternative to generic cursors by representing remote interactions using designs tailored to the visualization and interaction technique, thumbnail legends \replace{serve as a highly}{provide a} specific alternative to cursor legends. 
Relative to other representations, thumbnail legends allow \replace{the viewer to more directly see their}{more direct visibility into} collaborators' activity. However, they are less compact than cursor legends.

Thumbnails are designed as minimal representations of remote visualizations. Rather than simply scaling down a view, thumbnails strip out elements such as legends, titles, axes, and labels\,---\,information that is unnecessary for a high-level understanding of the remote interaction. Future work might explore ways to semantically simplify visualizations, for example, by selecting representative points in a scatterplot that communicate the overall shape of the data. Such visual reductions may require additional data transformation.

\section{Interactively Establishing Common Ground} \label{section:CollabInteractions}

Our design space describes how to design representations of collaborators that overlay or augment a visualization such that a user can conduct their own individual, independent exploration of the data.
However, \replace{by synchronizing only these representations, and not the entire visualization, we risk collaborators}{when independent explorations cause views to diverge, such as when two users zoom into different parts of the visualization, collaborators risk} losing common ground~\cite{clark1991grounding}.
To mitigate this risk, we present an additional layer of interactivity afforded by these cursors to allow for novel collaborative experiences. 

\subsection{Peeking}

We define \textit{peeking} as the action of temporarily setting the viewer's visualization state to that of a remote user. Semantically, peeking refers to a user's intention to glance at a remote user's view without necessarily committing their view to that state. We deem it important that users can easily peek at a remote user's visualization and then return to their original visualization state. In other words, peeking is a non-committal interaction between users, and the peeker does not relinquish control over their visualization at any given moment. By default, the viewer can peek at a remote interaction state by hovering over a cursor. The ephemeral quality of the mouse-over interaction matches well with the temporary nature of the peek interaction. Of course, peeking can be triggered by any low-level event the visualization designer sees fit, but the technique is important for providing users the ability to explore remote user states easily.

\subsection{Tracking}

We define \textit{tracking} as the action of actively synchronizing the viewer's visualization state to that of a remote user. In other words, when one user tracks another, their visualization mirrors the view of the remote visualization. This form of synchronization follows the work of most recent research collaborative visualization systems~\cite{polychrome, vistrates} in allowing users to view remote interactions in situ and in real-time. Tracking is a persistent version of peeking, and implementing one should effectively enable the other. In our implementation, the viewer can track other users by clicking on a cursor, which follows from hovering to peek at a remote interaction state.

\subsection{Forking}

We define \textit{forking} as the action of taking a remote user's interaction state as the starting point for new, divergent exploration. After a user has been tracking another user, they might identify something interesting in the remote view and choose to interact with the visualization themselves to modify the view.
Forking, by definition, ends the act of tracking another user, as new interactions modify local state separately from remote state. Forking is especially useful for visualizations that have many interaction techniques, such as several query filters. One might find a particular selection of filters interesting and choose to fork from that state by narrowing down on a subset of the data. Or, one might choose to flip one of the filters around to see how the same set of selections applies along a different dimension of the data.
Under our approach, as every user is giving their own instance of a collaborative visualization, forking occurs automatically when a user begins interacting with a visualization that was tracking another user.

%% file: sections/evaluation.tex
\section{\replace{Evaluation: Implementation in Vega-Lite}{Implementation and Example Gallery}}\label{chap:evaluation}

\begin{figure}
    \centering
    \includegraphics[width=\columnwidth]{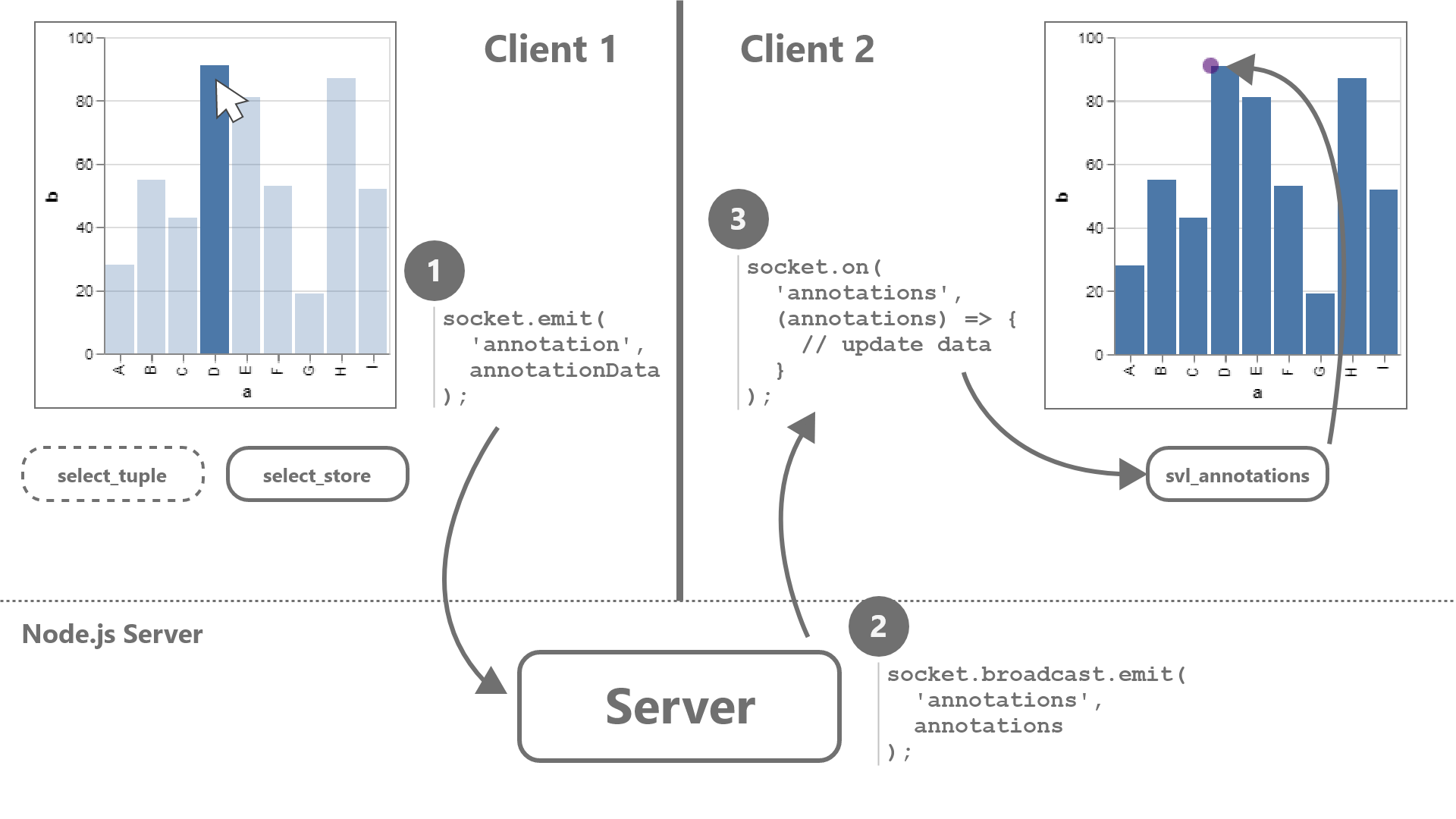}
    \vspace{-9mm}
    \caption{An overview of the how interactions on one client translate into cursors on another. (1) Client 1 reacts to a click event by emitting the corresponding data tuples to the server; (2) the server transmits this data to all connected clients; (3) Client 2 receives new data and updates its state, rendering new cursor marks on the visualization.}
    \vspace{-3mm}
    \label{fig:networking}
\end{figure}

We demonstrate the expressiveness of our designs by implementing a wrapper around the Vega-Lite visualization grammar~\cite{vegalite}. 
Our wrapper converts any Vega-Lite specification into a synchronous, multi-user collaborative visualization. 
Our interaction model consequently assumes interactions that are expressible as Vega-Lite selections, though future work may explore compatibility with custom interaction\del{ type}s. \del{The examples shown in \replace{Figure \ref{fig:gallery}}{\autoref{fig:gallery}} are implemented with our prototype, and demonstrate that collaborative visualization using our cursor designs and interaction techniques cover an expressive range of visualization types (histograms, scatter plots, line charts) and interaction types (brushing, query widgets, hover interactions, and single/multiple point selections).}%
Our implementation consists of \replace{two separate packages: a client and a server}{a client and a server package} (\autoref{fig:networking}). The client package \del{handles several tasks: it }annotates a given Vega-Lite specification with extra marks, signals, and data \replace{that renders}{to render} the cursors; it translates user interactions on the visualization into messages that are sent to \del{other }collaborators; and, it parses \add{incoming} messages \del{received from peers }to \replace{cause state changes to the visualization}{update visualization state}. The server package\del{ simply} relays information between \del{various }clients and maintains user state (e.g., color assignments).

\add{We evaluate expressiveness by creating an example gallery (\autoref{fig:gallery}). Examples are implemented with our prototype, demonstrating that our cursor designs and interaction techniques cover an expressive range of visualization types (histograms, scatter plots, line charts) and interaction types (brushing, query widgets, hover interactions, and single/multiple point selections).}

%% file: sections/conclusion.tex
\section{Conclusion}\label{chap:conclusion}

We present a design space for representing multiple users collaborating around a visualization without impeding an individual user's ability to read and analyze the visualization.
To re-establish common ground between collaborators, we additionally contribute \del{a set of }interactive techniques that allow users to peek or track their collaborators activity, or fork their own path forward.
We instantiate these ideas in a wrapper for \del{the }Vega-Lite\del{ visualization grammar}, and demonstrate coverage across a diverse range of visualization types and interaction techniques.
Our prototype implementation is available as open-source software at \url{https://github.com/mitvis/synchronized-vega-lite}.

Our design space and interaction techniques \replace{give us a foundation for exploring more}{enable further exploration of} complex \add{collaborative} user interactions\del{ across collaborative visualizations}. For example, Akia might be watching Deniz highlight \del{some }features of the\add{ir} data\del{ they have found in their research}. Akia might notice an interesting avenue of exploration and ask to take over\del{ control of the visualization}. Once Deniz relinquishes control, Akia \del{applies some }filters \del{to }the data as Deniz watches. Using our definitions, Akia started off tracking Deniz, then \del{chose to }fork\add{ed} \del{off of }Deniz's state, \replace{and}{as} Deniz switched \del{over }to tracking Akia. Perhaps this interaction \replace{could receive}{warrants} its own name, or \del{perhaps it }raises the question of when \replace{there should be multiple users controlling the same visualization at once}{explorations should diverge or come together}. \replace{Either way, the}{Our} core set of interaction techniques\del{ we have presented should} facilitate\add{s} further discussion about \del{the }user representation and collaboration on visualizations.


%% file: main.bbl
\begin{thebibliography}{10}

\bibitem{applegate1991technology}
L.~M. Applegate.
\newblock Technology support for cooperative work: A framework for studying
  introduction and assimilation in organizations.
\newblock {\em Journal of Organizational Computing and Electronic Commerce},
  1(1):11--39, 1991.

\bibitem{polychrome}
S.~K. Badam and N.~Elmqvist.
\newblock Polychrome: A cross-device framework for collaborative web
  visualization.
\newblock In {\em Proceedings of the Ninth ACM International Conference on
  Interactive Tabletops and Surfaces}, ITS ’14, p. 109–118. Association for
  Computing Machinery, New York, NY, USA, 2014. doi: {{%
10\hspace{.1pt}\discretionary{.}{%
}{.}\hspace{.4pt}1145\discretionary{/}{%
}{/}2669485\hspace{.1pt}\discretionary{.}{%
}{.}\hspace{.4pt}2669518}}


\bibitem{vistrates}
S.~K. {Badam}, A.~{Mathisen}, R.~{Rädle}, C.~N. {Klokmose}, and N.~{Elmqvist}.
\newblock Vistrates: A component model for ubiquitous analytics.
\newblock {\em IEEE Transactions on Visualization and Computer Graphics},
  25(1):586--596, 2019.

\bibitem{brodlie2004distributed}
K.~W. Brodlie, D.~A. Duce, J.~R. Gallop, J.~P. Walton, and J.~D. Wood.
\newblock Distributed and collaborative visualization.
\newblock In {\em Computer graphics forum}, vol.~23, pp. 223--251. Wiley Online
  Library, 2004.

\bibitem{clark1991grounding}
H.~H. Clark and S.~E. Brennan.
\newblock Grounding in communication.
\newblock 1991.

\bibitem{ubiquitous-analytics}
N.~Elmqvist and P.~Irani.
\newblock Ubiquitous analytics: Interacting with big data anywhere, anytime.
\newblock {\em Computer}, 46(04):86--89, apr 2013. doi: {{%
10\hspace{.1pt}\discretionary{.}{%
}{.}\hspace{.4pt}1109\discretionary{/}{%
}{/}MC\hspace{.1pt}\discretionary{.}{%
}{.}\hspace{.4pt}2013\hspace{.1pt}\discretionary{.}{%
}{.}\hspace{.4pt}147}}


\bibitem{grudin1994groupware}
J.~Grudin.
\newblock Groupware and social dynamics: Eight challenges for developers.
\newblock {\em Communications of the ACM}, 37(1):92--105, 1994.

\bibitem{hajizadeh2013supporting}
A.~H. {Hajizadeh}, M.~{Tory}, and R.~{Leung}.
\newblock Supporting awareness through collaborative brushing and linking of
  tabular data.
\newblock {\em IEEE Transactions on Visualization and Computer Graphics},
  19(12):2189--2197, 2013.

\bibitem{design-considerations}
J.~Heer and M.~Agrawala.
\newblock Design considerations for collaborative visual analytics.
\newblock {\em Information Visualization}, 7(1):49--62, 2008. doi: {{%
10\hspace{.1pt}\discretionary{.}{%
}{.}\hspace{.4pt}1057\discretionary{/}{%
}{/}palgrave\hspace{.1pt}\discretionary{.}{%
}{.}\hspace{.4pt}ivs\hspace{.1pt}\discretionary{.}{%
}{.}\hspace{.4pt}9500167}}


\bibitem{2008-graphical-histories}
J.~Heer, J.~Mackinlay, C.~Stolte, and M.~Agrawala.
\newblock Graphical histories for visualization: Supporting analysis,
  communication, and evaluation.
\newblock {\em IEEE Trans. Visualization \& Comp. Graphics (Proc. InfoVis)},
  14(6):1189--1196, 2008.

\bibitem{heer2012interactive}
J.~Heer and B.~Shneiderman.
\newblock Interactive dynamics for visual analysis.
\newblock {\em Queue}, 10(2):30--55, 2012.

\bibitem{2007-senseus}
J.~Heer, F.~B. Vi\'{e}gas, and M.~Wattenberg.
\newblock Voyagers and voyeurs: Supporting asynchronous collaborative
  information visualization.
\newblock In {\em ACM Human Factors in Computing Systems (CHI)}, pp.
  1029--1038, 2007.

\bibitem{hong2018collaborative}
S.~R. Hong, M.~M. Suh, N.~Henry~Riche, J.~Lee, J.~Kim, and M.~Zachry.
\newblock Collaborative dynamic queries: Supporting distributed small group
  decision-making.
\newblock In {\em Proceedings of the 2018 CHI Conference on Human Factors in
  Computing Systems}, CHI '18, p. 1–12. Association for Computing Machinery,
  New York, NY, USA, 2018. doi: {{%
10\hspace{.1pt}\discretionary{.}{%
}{.}\hspace{.4pt}1145\discretionary{/}{%
}{/}3173574\hspace{.1pt}\discretionary{.}{%
}{.}\hspace{.4pt}3173640}}


\bibitem{collab-vis-definition}
P.~Isenberg, N.~Elmqvist, J.~Scholtz, D.~Cernea, K.-L. Ma, and H.~Hagen.
\newblock Collaborative visualization: Definition, challenges, and research
  agenda.
\newblock {\em Information Visualization}, 10(4):310--326, 2011. doi: {{%
10\hspace{.1pt}\discretionary{.}{%
}{.}\hspace{.4pt}1177\discretionary{/}{%
}{/}1473871611412817}}


\bibitem{johansen1988groupware}
R.~Johansen.
\newblock {\em Groupware: Computer support for business teams}.
\newblock The Free Press, 1988.

\bibitem{2018-others-expectations}
Y.-S. Kim, K.~Reinecke, and J.~Hullman.
\newblock Data through others' eyes: The impact of visualizing others'
  expectations on visualization interpretation.
\newblock {\em IEEE Trans. Visualization \& Comp. Graphics (Proc. InfoVis)},
  2018.

\bibitem{mahyar_tory_2014}
N.~Mahyar and M.~Tory.
\newblock Supporting communication and coordination in collaborative
  sensemaking.
\newblock {\em IEEE Transactions on Visualization and Computer Graphics},
  20(12):1633–1642, 2014. doi: {{%
10\hspace{.1pt}\discretionary{.}{%
}{.}\hspace{.4pt}1109\discretionary{/}{%
}{/}tvcg\hspace{.1pt}\discretionary{.}{%
}{.}\hspace{.4pt}2014\hspace{.1pt}\discretionary{.}{%
}{.}\hspace{.4pt}2346573}}


\bibitem{2019-towards-visualive}
R.~Neogy, E.~Hu, and A.~Satyanarayan.
\newblock {Visualive: Representing Synchronized Visualization Interactions}.
\newblock In {\em IEEE InfoVis Posters}, 2019.

\bibitem{information-scent}
P.~Pirolli and S.~Card.
\newblock Information foraging.
\newblock {\em Psychological Review}, 106(4):643–675, 1999. doi: {{%
10\hspace{.1pt}\discretionary{.}{%
}{.}\hspace{.4pt}1037\discretionary{/}{%
}{/}0033\discretionary{%
}{-}{-}295x\hspace{.1pt}\discretionary{.}{%
}{.}\hspace{.4pt}106\hspace{.1pt}\discretionary{.}{%
}{.}\hspace{.4pt}4\hspace{.1pt}\discretionary{.}{%
}{.}\hspace{.4pt}643}}


\bibitem{vegalite}
A.~Satyanarayan, D.~Moritz, K.~Wongsuphasawat, and J.~Heer.
\newblock Vega-lite: A grammar of interactive graphics.
\newblock {\em IEEE transactions on visualization and computer graphics},
  23(1):341--350, 2016.

\bibitem{manyeyes}
F.~B. {Viegas}, M.~{Wattenberg}, F.~{van Ham}, J.~{Kriss}, and M.~{McKeon}.
\newblock Manyeyes: a site for visualization at internet scale.
\newblock {\em IEEE Transactions on Visualization and Computer Graphics},
  13(6):1121--1128, 2007.

\bibitem{wattenberg2005baby}
M.~Wattenberg.
\newblock Baby names, visualization, and social data analysis.
\newblock In {\em IEEE Symposium on Information Visualization, 2005. INFOVIS
  2005.}, pp. 1--7. IEEE, 2005.

\bibitem{collab-vis}
J.~{Wood}, H.~{Wright}, and K.~{Brodie}.
\newblock Collaborative visualization.
\newblock In {\em Proceedings. Visualization '97 (Cat. No. 97CB36155)}, pp.
  253--259, 1997.

\bibitem{wood1995cscv}
J.~Wood, H.~Wright, and K.~Brodlie.
\newblock Cscv-computer supported collaborative visualization.
\newblock In {\em Proceedings of BCS Displays Group International Conference on
  Visualization and Modelling}, pp. 13--25. Citeseer, 1995.

\bibitem{yi2007toward}
J.~S. Yi, Y.~ah~Kang, J.~Stasko, and J.~A. Jacko.
\newblock Toward a deeper understanding of the role of interaction in
  information visualization.
\newblock {\em IEEE transactions on visualization and computer graphics},
  13(6):1224--1231, 2007.

\end{thebibliography}
